\documentclass[aps,pre,10pt,notitlepage,showpacs,nofootinbib,superscriptaddress,twocolumn,raggedbottom,preprintnumbers,longbibliography]{revtex4-1}
\usepackage{amsfonts}
\usepackage{multirow}
\usepackage{amsmath}
\usepackage{amssymb}
\usepackage{array}
\usepackage{bbold}
\usepackage{bm}
\usepackage{booktabs}
\usepackage{dsfont}
\usepackage{enumitem}
\usepackage[perpage]{footmisc}
\usepackage{array}
\usepackage{float}
\usepackage{graphicx}
\usepackage[utf8]{inputenc}
\usepackage{lastpage}
\usepackage{nicefrac}
\usepackage[normalem]{ulem}
\usepackage[dvipsnames]{xcolor}
\usepackage{graphicx}
\usepackage{color}
\usepackage{bigstrut}
\usepackage{enumitem}
\usepackage{bm,mwe}
\usepackage[export]{adjustbox}
\usepackage{listings}
\usepackage{nameref}
\usepackage{dcolumn}
\usepackage{textcomp}
\usepackage{adjustbox}
\usepackage{soul}

\usepackage{hyperref}
\hypersetup{
    colorlinks=true,
    urlcolor=blue,
    linkcolor = blue,
    urlcolor  = blue,
    citecolor = blue,
    anchorcolor = blue}
\graphicspath{{./plots/}}

\newcolumntype{R}[2]{%
    >{\adjustbox{angle=#1,lap=\width-(#2)}\bgroup}%
    l%
    <{\egroup}%
}
\begin{document}

%%%%%%%%%%%%%%%%%%%%%%%%%%%%%%%%%%%%%%%%%%%%%%%
% \date{\today}

\title{
A Test of the Thermodynamics of Evolution
}

\author{Daniel Sadasivan}
\email{Daniel.Sadasivan@avemaria.edu}
\affiliation{Ave Maria University, Ave Maria, FL 34142, USA}

\author{Cole Cantu }
\email{cole.cantu@my.avemaria.edu}
\affiliation{Ave Maria University, Ave Maria, FL 34142, USA}

\author{Cecilia Marsh}
\email{cecilia.marsh@my.avemaria.edu}
\affiliation{Ave Maria University, Ave Maria, FL 34142, USA}

\author{Andrew Graham}
\email{andrew.graham@my.avemaria.edu}
\affiliation{Ave Maria University, Ave Maria, FL 34142, USA}
%

%%%%%%%%%%%%%%%%%%%%%%%%%%%%
%%%%%%%%%%%%%%%%%%%%%%%%%%%%
\preprint{}

%%%%%%%%%%%%%%%%%%%%%%%%%%%%%%%%%%%%%%%%%%%%%%%%%%%%%%%
%%%%%%%%%%%%%%%%%%%%%%%%%%%%%%%%%%%%%%%%%%%%%%%%%%%%%%%
\begin{abstract}
Recent research has extended methods from the fields of thermodynamics and statistical mechanics into other disciplines. Most notably, one recent work creates a unified theoretical framework to understand evolutionary biology, machine learning, and thermodynamics. We present simulations of biological evolution used to test this framework. The test simulates organisms whose behavior is determined by specific parameters that play the role of genes. These genes are passed on to new simulated organisms with the capacity to mutate, allowing adaption of the organisms to the environment. With this simulation, we are able to test the the framework in question. The results of our simulation are consistent with the work being tested, providing evidence for it. 

 \end{abstract}
%%%%%%%%%%%%%%%%%%%%%%%%%%%%%%%%%%%%%%%%%%%%%%%%%%%%%%%
%%%%%%%%%%%%%%%%%%%%%%%%%%%%%%%%%%%%%%%%%%%%%%%%%%%%%%%

\maketitle

%%%%%%%%%%%%%%%%%%%%%%%%%%%%%%%%%%%%%%%%%%%%%%%%%%%%%%%
%%%%%%%%%%%%%%%%%%%%%%%%%%%%%%%%%%%%%%%%%%%%%%%%%%%%%%%

\section{Introduction}

\subsection{Background}
\label{subsec:Background}
%%%%%%%%%%%%%%%%%%%%%%%%%%%%%%%%%%%%%%%%%%%%%%%%%%%%%%%
%%%%%%%%%%%%%%%%%%%%%%%%%%%%%%%%%%%%%%%%%%%%%%%%%%%%%%%

Thermodynamics and statistical mechanics have made significant progress in describing the emergent behavior of particles. Similarly, the study of evolutionary biology also requires an understanding of the rates of emergent macroscopic properties in a very large system. There has been a longstanding effort to develop a rigorous quantitative framework to describe rates of processes in evolution. A large body of  research has been devoted to this goal, both from previous decades~\cite{fisher1999genetical,10.1093/genetics/16.2.97,Kimura1974}, as well as from more recent work~\cite{Lynch2016GeneticDS,muraleedharan2024geometric,mayer2023order,summers2023entropic,Aristov,Aristov2,Aristov3,Duan,Persi,Persi2,Smerlak,Frank,FrankPrice}. Notably,  equilibria and phase-transition-like changes in evolution have been studied~\cite{Bakhtin,Mathis}. The work has been extended to biological systems in other contexts as well~\cite{Tomkins,PhysicalFoundationsOfBiologicalComplexity}.
In addition, research has made progress relating the framework of evolution and other complex biological systems to issues in physics~\cite{Sotnikov2023EmergenceOC}, neuroscience,~\cite{friston2023variational,FriedmanAnts,KozyrevAndPechen}, and other areas. 

Some research has focused on relating thermodynamics and evolution directly~\cite{Igamberdiev2023TowardTR,PhysicalPrinciplesOfEvolution}. One of the most cited works in this field is Ref.~\cite{SellaHirsh}, expanded upon by Ref.~\cite{Barton2009OnTA}.These papers present a model which adapts a statistical mechanics formalism to evolution. In Ref.~\cite{SellaHirsh}, multiple parallels are drawn between thermodynamic variables and evolutionary variable, the most relevant to this work being their identification of population size as the evolutionary equivalent of temperature. The work being tested here, Ref.~\cite{ToE}, makes a contrary claim about the nature of evolutionary temperature, instead identifying it with the stochasticity of an evolutionary system.

The attempt to develop a mathematical framework for rates of evolution is part of a broader effort to understand and describe thermodynamic systems out of equilibrium. There have been many recent advances in these fields, especially in terms of spin glasses~\cite{SpinGlassBook,SpinGlassesEdwards,SpinGlassesBinder,Yuan}, and the Mpemba effect~\cite{Lu_2017,Amorim_2023,PhysRevX.9.021060}.

The field of machine learning also studies the emergent properties of a large number of individual components. One example is the well-known evolutionary algorithm. ~\cite{yu2010introduction,Sui} Many further parallels can be drawn between biological evolution and neural networks, as both systems are constantly "learning" and adapting to their environments. Because of these similarities, some research has worked to develop a theory unifying these fields~\cite{kozyrev2023learning, Sinitskiy2023MakingSO} and apply advances from one field to another~\cite{GrabovskyVanchurin,CohenMarron,ChrisLU, andrejic}.

Additional work has been done relating physical and chemical systems to neural networks~\cite{hong2022melting,Sinitsky,KatsnelsonQuantum,TheWorldAsANeuralNetwork} as well as towards a general framework for neural networks~\cite{Hochberg,TowardsATheoryofMachineLearning}.

This paper tests the work of Ref.~\cite{ToE},Thermodynamics of Evolution and the Origin of Life, and ~\cite{Vanchurin_2022}, Towards a Theory of Evolution as Multilevel Learning, which relate the principles of thermodynamics, machine learning, and biological evolution as optimization processes. This work has been applied to the evolution and spread of SARS-CoV-2 in in Ref.~\cite{RomanenkoVanchurin}. 

\subsection{Thermodynamics of Evolution}
\label{subsec:ToE}
This section summarizes Ref.~\cite{ToE}, which gives a framework for unifying thermodynamics, evolution, and machine learning. The article makes the case that this framework could be used to understand the origin of life. Our simulations only test the Thermodynamics of Evolution, and do not test applications to the origin of life. As such, we will only summarize the Thermodynamics of Evolution in this section.

 The major principle of Refs.~\cite{ToE} is the minimization of entropy subject to constraints. Entropy is identified as
\begin{align}
S=-\int d^N \textbf{x} \hspace{1 mm} p(\textbf{x}|\textbf{q}) \log \left(p(\textbf{x}|\textbf{q})\right),
\label{eq:VanchurinEntropy}
\end{align}
where $\textbf{x}$ represent non-trainable variables describing the environment and $\textbf{q}$ representing trainable variables that adapt to the environment through the evolution process.  $p(\textbf{x}|\textbf{q})$ is the probability distribution of $\textbf{x}$ given $\textbf{q}$.

The minimization of entropy is subject to two constraints: first, that the integral over all probabilities is equal to 1, and second, that an evolutionary quantity called the additive fitness, $H$, (which is made analogous to energy in thermodynamics), can be averaged with 
\begin{align}
U(\textbf{q})=\int d^N x \hspace{1 mm} H(\textbf{x},\textbf{q}) p(\textbf{x}|\textbf{q}),
\label{eq:VanchurinU}
\end{align}
where $ U(\textbf{q})$ does not depend on $ p(\textbf{x}|\textbf{q})$. The entropy is minimized subject to these constraints from which the probability is derived:
\begin{align}
p(\textbf{x}|\textbf{q})=\frac{e^{-\beta H(x,\textbf{q})}}{Z(\beta,\textbf{q})},
\label{eq:VanchurinProbability}
\end{align}
where, $\beta $ is the Lagrange Multiplier for the constraint $U$.
The symbol $\beta$ is chosen because it is identified as being equivalent to $\beta$ in thermodynamics which is the inverse temperature. In the context of evolution, the temperature is the "extent of the stochasticity of the system." 

$Z$, the evolutionary partition function in Eq.~\eqref{eq:VanchurinProbability}, is given by
\begin{align}
Z(\beta|\textbf{q})=\int d^N x \hspace{1 mm} \phi(\textbf{x},\textbf{q}),
\label{eq:VanchurinPartitionFunction}
\end{align}
where $\phi$ is identified as the Malthusian Fitness. Just as in thermodynamics and statistical mechanics, the quantity $Z$ contains information about many other quantities. One such quantity is the free energy, $F$, defined as
\begin{align}
F\equiv -T \log Z.
\label{VanchurinFreeEnergy}
\end{align}

Ref.~\cite{ToE} proposes an empirical test within the context of an ideal mutation model. Like the ideal gas model, the ideal mutation model makes assumptions to simplify calculations. The test in the ideal mutation model assumes the following: both the population size and number of genes are sufficiently large, there is no epistasis (genes cannot suppress the effects of other genes), the population remains fixed, and, finally, that most mutations are not beneficial, but beneficial mutations take a negligible amount of time to spread to the rest of the population.  

From these assumptions, the work derives the following equation for the partition functions of a system that evolves from temperature 1 to temperature 2, $Z^{(1)}$ and $Z^{(2)}$.
\begin{align}
\frac{\log Z^{(1)}(\textbf{q})}{\log Z^{(2)}(\textbf{q})}=\frac{\beta_1 F(\textbf{q})}{\beta_2 F(\textbf{q})}=\frac{\beta_1}{\beta_2}=\frac{T_2}{T_1}
\label{eq:VanchurinTestEquation}
\end{align}
It follows from this equation that the ratio of the initial and final partition functions does not depend on the initial or final genes. This concrete prediction can be tested. We test a modified version of this equation described in Sec.~\ref{sec:Methods}.

Additionally, the work accounts for an evolutionary potential, $\mu$, which is the amount of evolutionary work needed to add an additional gene, however, it is irrelevant to our work, in which the number of genes is held constant. 
 
\section{Methods}
\label{sec:Methods}

We created a simulation to test the claims of Ref.~\cite{ToE}. This simulation tracks "organisms" whose behavior is controlled by a set of "genes." These  organisms are capable of reproduction when certain conditions are met. When reproducing, the resulting organism will have a mutated version of the "parent" organism's genes. The evolution process adjusts these genes so the organisms are more capable of survival and reproduction in their environment. As a result, this simulation is an ideal test case because it can test predictions while being able to track all existing information about the system. This simulation is described qualitatively in the main body of the text with technical, numerical details described in Appendix~\ref{sec:SimulationDetails}.

\begin{figure}
\includegraphics[width=0.99\linewidth]{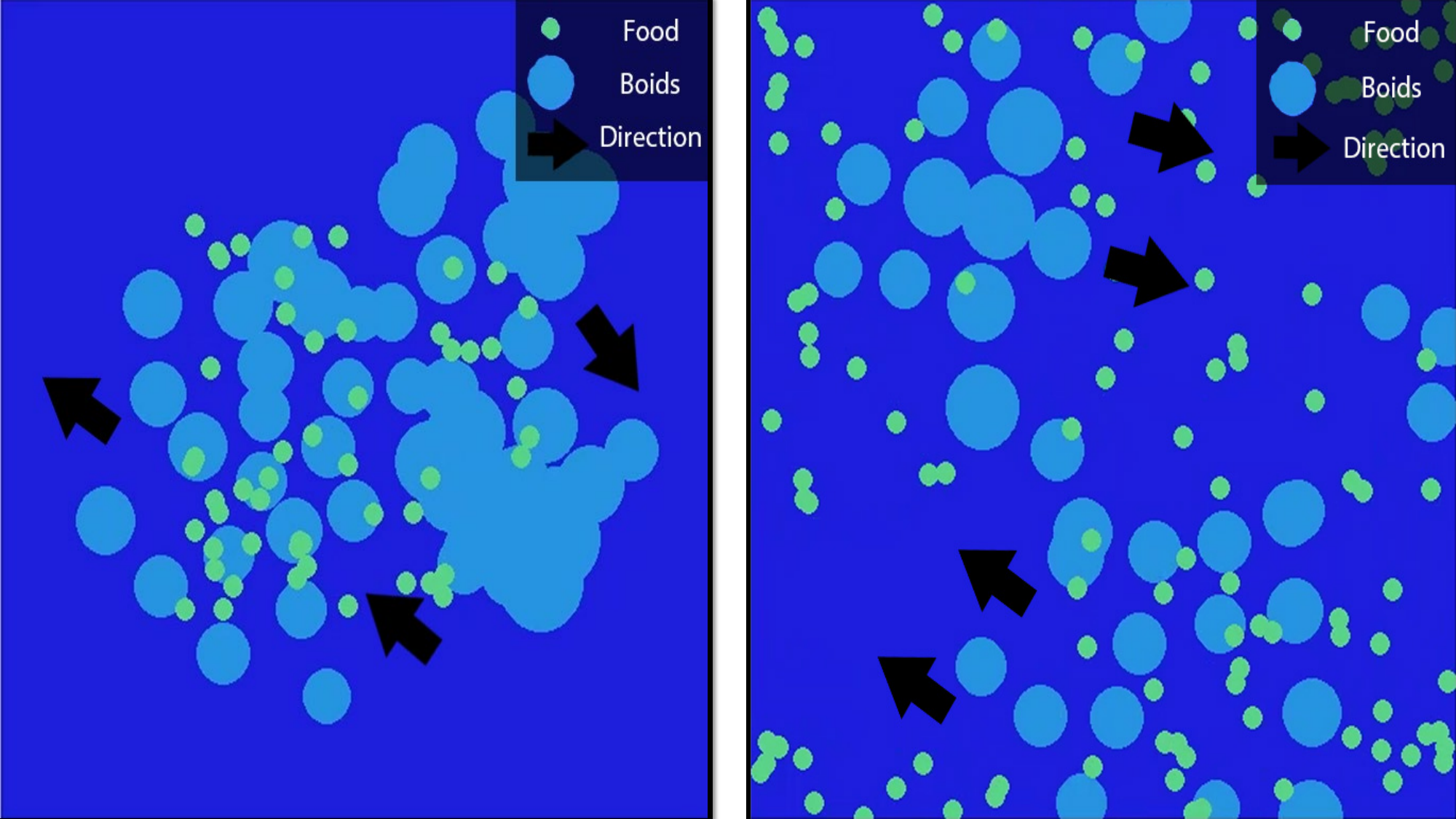}
\caption{Visualization of the simulation. Large circles represent the organisms, whose motion is controlled by parameters that are analogous to genes. Small dots represent food. Arrows show 
 the general movement of the organisms. Over time, the mutation of "genes" during reproduction adapts the organisms to the sources of food. Generating the food in a smaller area as shown in the figure on the right changes the evolutionary equivalent of temperature.}
\label{fig:Illustration}
\end{figure}
Each organism accelerates in proportion to its net force, the sum of all the component "forces" on it.  These forces are designed to create an emergent system, that is, an environment which cannot be understood by looking at an individual organism without performing the complete calculation. We have created a system where the forces governing the behaviors follow the rules for Boid motion, employed in a number of works, including Refs. ~\cite{drones7110673,greig2023,refId0,Hartman06CAVW}. A pedagogical explanation with pseudocode is given in Ref.~\cite{Boids}. Along with the forces from Boid motion, the organisms experience a random force. The relative strengths of the forces from the Boid motion and the random force are determined by the genes of a given organism. 

The principles of Boid motion have been chosen for this simulation because of their ability to create an emergent environment. The purpose of this simulation was to provide a complex testing environment. If the environment had been too simple, the results could have been predicted with statistical methods such as Brownian motion. If the simulation's results could be exactly calculated without running, it, the methods Ref.~\cite{ToE} presents would provide no insights.

Each organism has an "energy" value. If the energy of a given organism decreases to below 0, the organism is removed from the simulation. This energy decreases when the organism moves or reproduces, and increases with the consumption of "food". Units of food, (which can be thought of as producers), are generated at random points in a portion of the simulation. 

A visualization of the motion of each organism is shown in Fig.~\ref{fig:Illustration}. If an organism passes over a unit of food, the food will be consumed and the organism will gain the food's energy. The area covered by an organism depends on its size, which is determined by genes. If one organism moves over another one, it might be able to gain energy by consuming the smaller organism. The greater the size of the consuming organism, the higher the likelihood of it consuming the smaller one. The smaller organism is removed from the simulation once it has been consumed. The specific rules for the amounts of energy gained or lost during these interactions are described in the appendix.

As mentioned in Sec.~\ref{subsec:ToE}, the evolutionary temperature is identified as the stochasticity of the system. This stochasticity should not be confused with amount of randomness involved in individual steps in the code. Sometimes, a simulation in which organisms have many randomly determined behaviors can lead to very precisely determined, non-random result, because the random choices average out. Such a simulation would have a low temperature not a high one. Instead, the stochasticity in this context should be understood as the unpredictability of the macrostate.

To vary the temperature in our simulation, we vary the area in which food is generated.  This can be seen in Fig.~\ref{fig:Illustration},where the left side shows a case where food is generated in a smaller central region and the right side shows a case where food is generated throughout the area.  We have chosen 5 temperatures of various food areas, given in Tab~\ref{tab:NumbersOfSimulations}. The area of food generation corresponding to minimum stochasticity is neither the largest nor the smallest area. For the smallest 2 of the 5 chosen values, the area is small enough that there is a significant chance that no organism with fit genes will reproduce in the food-generation area in which their offspring will likely survive. A large stochasticity results from the large gap in fitness between the case where organisms do reproduce in the food area and the case where they do not. On the other hand, in the largest 2 of the food generation areas, there is greater uncertainty about where each food will be generated. Organisms can adapt less readily to less predictable environment. The genes of organisms can less effectively be adapted to consume this food, resulting in more cases where the organisms quickly die off. This uncertainty about the growth rate also results in a larger stochasticity. Thus, the middle food area results in the lowest temperature.

The simulation employed in this work does not exactly match the ideal mutation model presented in Ref.~\cite{ToE}. There are several notable differences, including that, in the ideal mutation model, the population does not change whereas in our simulation, it does. We have differed from the ideal mutation model to allow for the calculation of the partition function using Eq.~\eqref{eq:VanchurinPartitionFunction}. 
The Malthusian fitness, $\phi$, is defined for evolutionary biology in Ref.~\cite{Vanchurin_2022} as "the rate of change of the prevalence of the given genotype in an evolving population." In order for $\phi$ to be nonzero, either the population under study must change or other populations must change. Thus, the population of this simulation is allowed to change. Additionally, the genes in the ideal mutation model remain unchanged until a relevant mutation occurs. When this mutation occurs, it proceeds quickly to fixation. In our simulation, we have chosen genes to create an emergent environment.  If we were to impose this condition in our simulation, we would be altering the environment. Thus, there is a non-negligible change in genes as the organisms evolve. Furthermore, because we are simulating evolution of organisms with asexual reproduction, the time to fixation is longer because genes cannot be spread as quickly through a population as they can in organisms that reproduced sexually~\cite{Melian2012}.  The differences between this simulation and the ideal mutation model should be understood as small variations left after attempting to reach as close as possible to the idealization.

Because of the distinctions between our simulation and the ideal mutation model, the proposed test from Ref. ~\cite{ToE}, given in Eq.~\eqref{eq:VanchurinTestEquation}, can no longer be taken as an exact equality. Because of this, we perform a very similar experiment to the one proposed in the original work. Instead of taking a population at a temperature, $T_1$, and letting it evolve to temperature $T_2$ and then comparing  $\frac{\log{Z^{(1)}}}{\log{Z^{(2)}}}$ to $\frac{T_2}{T_1}$, we choose a starting set of genes, $\textbf{q}_i$ and put the organisms with these genes into environments of different stochasticities, corresponding to different temperatures. Over a short amount of time, $\textbf{q}_i$ will change slightly in response to its environment. We can call the new genes $\textbf{q}_f^T$, where the final set  of genes will depend on the temperature as well as the initial genes. If the simulation is only run for a short time, then $\textbf{q}_f^T=\textbf{q}_i+\Delta \textbf{q}^T$, where $\Delta \textbf{q}^T$ is small. We then take the ratios of the logs of the partition functions for $\textbf{q}_f^{T_1}$ and $\textbf{q}_f^{T_2}$. Thus, the test condition, Eq.~\eqref{eq:VanchurinTestEquation}, modified for this simulation becomes

\begin{align}
\nonumber \frac{\log Z^{(1)}(\textbf{q}_f^{T_1})}{\log Z^{(2)}(\textbf{q}_f^{T_2})}=\frac{\beta_1 F(\textbf{q}_f^{T_1})}{\beta_2 F(\textbf{q}_f^{T_2})} =\frac{T_2 F(\textbf{q}_i+\Delta \textbf{q}^{T_1})}{T_1 F(\textbf{q}_i+\Delta \textbf{q}^{T_2})} \\
=\frac{T_2 F(\textbf{q}_i)}{T_1 F(\textbf{q}_i)} +\mathcal{O}(\Delta \textbf{q}^{T_1}) +\mathcal{O}(\Delta \textbf{q}^{T_2}) \approx \frac{T_2}{T_1}.
\label{eq:ActualTestEquation}
\end{align}
In the last equals sign, we have done a Taylor expansion of the term, remembering that $\Delta \textbf{q}_f^T$ is small because we have run the simulation only for a short time.

The simulation was run for 6 different starting sets of genes, $\textbf{q}$. The values for $q_i$ are given in Tab.~\ref{tab:genetable}. These genes were chosen by the following process: A list of random genes is generated. Then, the simulation is run once for each entry in this list with the organisms having the starting genes of the entry in the list. Genes which led to the organisms immediately dying were removed from the list. Once this process was complete, 6 sets of genes were randomly selected from the list. This process was used to decrease the likelihood that the starting population would  immediately die out. Most randomly generated genes lead to a high probability of all organisms dying before reproduction. The process we have used does reduces the probability of immediate extinction although it does not eliminate the possibility.

For each of the six genes, we repeatedly use them as the starting genes for 30 organisms for each of 5 different temperatures. Tab.~\ref{tab:NumbersOfSimulations} gives the number of times the simulation was run for each gene and area of food which determines the evolutionary temperature. 

\begin{table}
\begin{tabular}{|l | c | c | c | c | c | c | c |}
\hline
 & $\textbf{q}_1$ & $\textbf{q}_2$ & $\textbf{q}_3$ & $\textbf{q}_4$ & $\textbf{q}_5$ & $\textbf{q}_6$ & Food Area \\
 \hline
 $T_1$ & 174 & 174 & 174 & 174 & 174 & 174 & 1.00 \\
  $T_2$ & 133 & 133 & 133 & 133 & 133 & 133 & 0.75\\
   $T_3$  & 142 & 142 & 142 & 141 & 141 & 141 & 0.52\\
    $T_4$  & 167 & 167 & 167 & 166 & 166 & 166 & 0.19\\
     $T_5$  & 108 & 108 & 108  & 108 & 108 & 108 & 0.07\\
\hline

\end{tabular}
 \caption{The number of times the simulations was run for each $\textbf{q}$ and $T$.The right column gives the area in which food is generated as a fraction of the total area. Varying this area varies the evolutionary temperature.}
 \label{tab:NumbersOfSimulations}
\end{table}

%%%%%%%%%%%%%%%%%%%%%%%%%%%%%%%%%%%%%%%%%%%%%%%%%%%%%%%%%%%%%%%%%%%%%%%%
%%%%%%%%%%%%%%%%%%%%%%%%%%%%%%%%%%%%%%%%%%%%%%%%%%%%%%%%%
\section{Results}
\label{sec:Results}

Repeatedly running the simulation provides results which are used to test Eq.~\eqref{eq:ActualTestEquation}.  Because the approximate equals sign only holds for small changes in genes, $\Delta\textbf{q}^T$, there is a narrow range of times over which the simulation ought to be analyzed. If the simulation is run for too long a time, the genes will change considerably and Eq.~\eqref{eq:ActualTestEquation} ceases to be a valid approximation. However, if the simulation is terminated after too short a time, certain fit combinations of genes that might eventually lead to population growth have not done so. In this situation, the true fitness of these genes will not be measured. The method employed here is to run the simulation and stop it once the population size has reached a local maximum. This ensures that the simulation will account for the fitness of the genes, but prevents the simulation from running too long.

The partition function from Eq.~\ref{eq:ActualTestEquation} can be calculated from the Malthusian fitness, $\phi$ following Eq.~\eqref{eq:VanchurinPartitionFunction}. The Malthusian fitness is defined in Ref.~\cite{Vanchurin_2022} as "the expected reproductive success of a given genotype: that is, the rate of change of the prevalence of the given genotype in an evolving population," following the work of Ref.~\cite{crow2017introduction}. To make this definition quantitative, we calculate the Malthusian fitness for a given trial number $\ell$ with 

\begin{align}
\label{eq:phi}
\phi_\ell=e^{\bar{n}_\ell},
\end{align}
where, $\bar{n}$ is the average number of offspring each organism has. The calculation of the this number is complicated by the question of how to treat organisms that survive at the end of the simulation. These organisms would continue to have offspring if the simulation were run longer. We do not count these organism towards the average. This is effectively the same as assuming that the average number of offspring of surviving organisms will be the same as the average number of offspring throughout the simulation. Thus, the average number of offspring for a simulation that has reached its peak could be written 
\begin{align}
\label{eq:averagenumberofoffspring}
\bar{n}=\frac{n_t}{n_t+n_i-n_s},
\end{align}
 where $n_t$ is the total number of organisms to be created during the simulation, $n_s$ is the number of organisms surviving at the peak population of the simulation, and $n_i$ is the number of organisms at the start of the simulation. 
 
 Note that if the population only declines, the peak will occur at the start of the simulation. If Eq.~\eqref{eq:averagenumberofoffspring} is evaluated at the start of the simulation, $n_t+n_i-n_s=0$ and $\bar{n}$ becomes undefined. To prevent these scenarios, $\bar{n}$ is set to 0 if the peak occurs in the first five time steps.

With the Malthusian fitness, the partition function can be calculated by integrating over all variables describing the environment, as stated in Eq.~\eqref{eq:VanchurinPartitionFunction}. The environment employed in this work involves generating food at random locations and letting the organisms adapt to the environment. Numerically, this integral is done using Monte Carlo methods, by summing over the randomly generated environments. Thus, the partition function is calculated with
\begin{align}
\label{eq:SadasivanPartitionFunction}
Z=\frac{1}{N}\sum_{\ell=1}^N \phi_\ell,
\end{align}
where $N$ is the number of trials. As mentioned in the previous section, the number of trials for each set of genes and temperature is given in Tab.~\ref{tab:NumbersOfSimulations}.

\begin{figure*}[t]
\includegraphics[trim=80 50 80 80,width=0.8\linewidth]{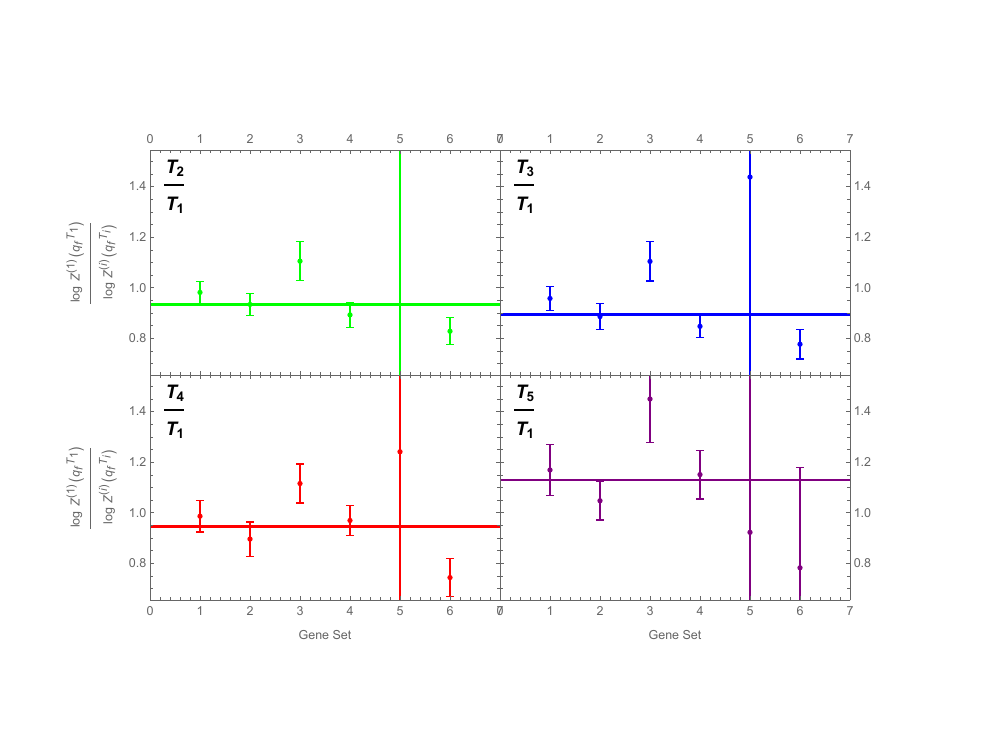}
\caption{Plot of ratios of temperatures with respect to each set of genes. One-standard-deviation uncertainties are shown. If predictions are correct, each temperature ratio will be approximately independent of the starting gene set. The horizontal lines correspond to values of $x$ that return minimum $\chi^2(x)$. This values gives the best prediction for the ratio $T_j/T_1$. }
\label{fig:TemperatureRatios}
\end{figure*}

The ratios of logs of partition functions (used for Eq.~\eqref{eq:ActualTestEquation}) are shown with their uncertainties in Fig.~\ref{fig:TemperatureRatios}. These uncertainties are calculated using methods described in a standard metrology or mathematical statistics textbooks such as Ref.~\cite{freund1971mathematical}. We first show our calculation of the uncertainty for the partition function for a given set of starting genes and temperature, $u_{Zq_i T_j}$. There are two contributions to this uncertainty. Firstly, there is the random uncertainty coming from the standard deviation of the partition function, $u_{rq_iT_j}$. This uncertainty is 
\begin{align}
u_{rq_iT_j}=\frac{\sigma_{{q_iT_j}}}{\sqrt{N_{q_iT_1}}}
\end{align}
where $\sigma_{{q_iT_j}}$ is the standard deviation of $\phi$ for the given temperature and genes and  $N_{q_iT_i}$ is the number of times that the simulation was run for the given $q$ and $T$. This number is given in Tab.~\ref{tab:NumbersOfSimulations}.
In addition to this uncertainty, there is the uncertainty coming from the stopping point for our calculations for the simulation. As discussed previously,  the Malthusian fitness is calculated over the range of time from the beginning of the simulation until the population reaches a local maximum. This decision is made to prevent the simulation from running for too large or small a time, however, the choice of what range of time over which to evaluate the simulation is somewhat arbitrary. Because of this, we introduce an uncertainty associated with the time over which the simulation was run. This uncertainty, $u_c$ is smaller than $u_r$, but it is non-negligible. We calculate it with the following formula
\begin{align}
u_{cq_iT_j}=\frac{1}{N_{q_iT_j}}\sqrt{\sum_{k=1}^N \sigma_{cq_iT_j}^2 },
\end{align}
where $\sigma_{cq_iT_j}$ is the standard deviation of three values of $\phi$, one calculated with $\phi$ when the population is at the local maximum for population, one calculated one time step before that, and one calculated two time steps before that. In most cases, this standard deviation is very close to 0 because $\phi$ is not very sensitive to the exact time at which it is calculated, however, for some $\phi$ it is sensitive enough to provide a small but non-negligible uncertainty. 

The combined uncertainty for these two sources is 
\begin{align}
u_{Zq_iT_j}=\sqrt{u_{rq_iT_j}^2+u_{cq_iT_j}^2}
\end{align}
The total uncertainty for ratio of log which is plotted in Fig.~\ref{fig:TemperatureRatios} is given by
\begin{align}
u_{r}=\sqrt{\left|u_{Zq_iT_1}\frac{\partial}{\partial Z_1} \left(\frac{\log{Z_1}}{\log{Z_2}}\right)\right|^2  +\left|u_{Zq_iT_2}\frac{\partial}{\partial Z_2} \left(\frac{\log{Z_1}}{\log{Z_2}}\right)\right|^2}.
\end{align}
These one-standard-deviation uncertainties are shown in Fig.~\ref{fig:TemperatureRatios}. If the test equation, Eq.~\eqref{eq:ActualTestEquation} is satisfied, the ratios will not depend strongly on $\textbf{q}$. This means that each of the ratios $T_i/T_1$ should be similar regardless of which $q$ is used. 

We start by finding the most probable value for the ratio, $T_i/T_1$, if it is constant. This is done by minimizing a $\chi^2(x)$ value. 
\begin{align}
 \chi^2(x)=\sum_{i=1}^n R_i^2(x), \hspace{2 mm} R_i(x)=\frac{x-T_{2q_i}/T_{1q_i}}{u_{ri}},
 \label{eq:chisquared}
\end{align}
where $n$ is the number of data points and $R_i$ are known as the residuals. The value of $x$ that gives the smallest $\chi^2(x)$ is the best prediction for $T_i/T_1$.

This estimate is used rather than taking the average of $T_i/T_1$ for each $\textbf{q}$, because the different values have different uncertainties. The $\chi^2(x)$ minimization weights each point according to the confidence from the uncertainties. We can then test how likely it is that, if this is the true value, the ratios with the given uncertainties and values would result. From the $\chi^2(x)$, we can find the uncertainties on the temperature ratios, $u_{T_i/T_1}$, and the $p$-value. A low $p$-value indicates that it is unlikely that the ratios are independent of the starting genes. Note that the $p$-values are calculated neglecting the $\mathcal{O}(\Delta \textbf{q}^{T})$ terms in Eq.~\eqref{eq:ActualTestEquation}. In other words, they are the probability of obtaining this result if the calculated ratio were the true and if Eq.~\eqref{eq:ActualTestEquation} were an exact equality. The ratios $T_j/T_1$, the minimum $\chi^2(x)$ values, and the $p$-values are given in Tab.~\ref{tab:Pvaluesandthings}.
\begin{table}
\begin{tabular}{| l | c | c | c | c |}
 \hline
 & $T_2$ & $T_3$ & $T_4$ & $T_5$\\
 \hline
 $T_i/T_1$ & 0.93471 & 0.895286 & 0.945315 & 1.13011 \\
 \hline
  $u_{T_i/T_1}$ & 0.0226957 & 0.0238207 & 0.0300963 & 0.0484969 \\
 \hline
 $\chi^2(x)$ & 10.8206 & 14.3142 & 13.2294 & 5.60171 \\
  \hline
 $p$-value & 0.0550546 & 0.0137319 & 0.0213212 & 0.346922 \\
  \hline
\end{tabular}
\caption{The temperature ratios, $T_i/T_1$ with corresponding uncertainties $u_{T_i/T_1}$, $\chi^2$ values, and \textit{p}-values for each temperature. Calculation for each of these quantities is described in the text.}
\label{tab:Pvaluesandthings}
\end{table}
\begin{figure}
\includegraphics[width=0.8\linewidth]{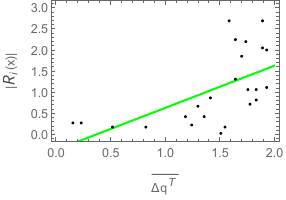}
\caption{Plot of the absolute value of the residuals, $|R_i(x)|$ of each temperature ratio and set of starting genes with respect to the average change in genes, 
$\overline{\Delta \textbf{q}^T}$.
Even though the data are non-linear, a best fit line is shown to demonstrate that the relationship between the variables is even stronger than the correlation coefficient indicates.  }
\label{fig:QvsR}
\end{figure}

As previously argued, Eq.~\eqref{eq:ActualTestEquation} is not an exact equality. If the equation is true, one would expect that the ratio of logs calculated for a temperatures and starting genes in the instances where the average genes change substantially, there would be a greater deviation from the average value, (which can be quantified with the residual, $R_i$ from Eq.~\eqref{eq:chisquared}) than in cases without substantial change in genes. Fig.~\ref{fig:QvsR} shows data testing this expectation. This figure plots the residual, $R_i$ for each of the 24 data points in Fig.~\ref{fig:TemperatureRatios} vs. the the average change in genes. The average, percentage change in genes for a given trial is calculated with 
\begin{align}
\Delta \textbf{q}^T=
\sum_{j=1}^{n_g}\frac{2|{q_{f}^T}_j - {q_i^T}_j|}{|{q_{f}^T}_j + {q_i^T}_j|},
\end{align}
where ${q_i^T}_j$ is the initial value for gene $j$ for temperature $T$ in a starting set of genes and  ${q_f^T}_j$ is the  final value for gene $j$ at temperature $T$ that evolved from the starting set of genes. If this quantity is averaged over all trials, it is denoted with an overline: $\overline{\Delta \textbf{q}^T}$

\section{Analysis}
\label{sec:Analysis}

In this section, the results from the previous section are interpreted. While the results are not conclusive, they are entirely consistent with the claims of Ref.~\cite{ToE}. These tests could have proven this reference false and did not, giving evidence in favor of the work.

The $p$-values values in Tab.~\ref{tab:Pvaluesandthings} (which quantify how likely the data points in Fig.~\ref{fig:TemperatureRatios} match the lines) are small enough that it is difficult to reasonably conclude that Eq.~\eqref{eq:VanchurinTestEquation} is true as an exact equality. Two of the four are above the 
 more restrictive, traditional 5\% threshold and two of the four are not. However, of the two that are not, both are above the more permissive traditional 1\% threshold. From this we can conclude that it is not reasonable to conclude that $\frac{\log{Z_1}}{\log{Z_2}}$ is completely independent of $\textbf{q}$, however, if the values for each ratio were slightly different, it would be independent of Temperature. This is exactly what would be predicted by Eq.~\eqref{eq:ActualTestEquation}. 

Further confirmation of this equation is given in Fig.~\ref{fig:QvsR}. The values of $R_i$ show a clear dependence on $\Delta\textbf{q}$. 
There is a strong correlation coefficient of 0.598. It should be noted that a correlation coefficient is a measure of how closely the data follow a line. Because the data visibly follow a pattern that is not a line, there is an even stronger relationship between the two quantities than would be expected by the correlation coefficient. There is no reasonable chance that these two quantities are unrelated. 

This is noteworthy because  the smaller $\Delta \textbf{q}$ is, the closer Eq.~\eqref{eq:ActualTestEquation} is to an exact equality. One would expect that if Eq.~\eqref{eq:ActualTestEquation} is true,  the $p$-values in Tab.~\ref{tab:Pvaluesandthings} would be close to the cutoff threshold because some of the cases where $\Delta \textbf{q}$ is large, the residuals contributing to $\chi^2(x)$, (which are the source of the small $p$-value) are larger than they would be if ~\eqref{eq:ActualTestEquation} were an exact equality. Thus, the cases with large change in genes would also have large residuals. This is exactly what is observed, which provides a strong confirmation of the theory.

The data in this research are consistent with the claims that are being tested. However, this consistency does not necessarily prove that any other explanations are inconsistent with the data observed. The data still have relatively large uncertainties, which could mean that there are alternative explanations for the results. We investigate two alternative explanations and find that neither of them can account for the results. 

One explanation is that the values of  $\log{Z_T}$, in the limit of infinite data, would all be so similar to each other for each temperature that the ratios of these logs approaches 1. If this were the case, any deviation of the ratio of logs from 1 in the present data is merely a statistical fluctuation. This explanation can be tested quantitatively. The $\chi^2(1)$ distributions is used for each of the four ratios to calculate probability of observing data as extreme as those plotted. These probabilities are respectively: 0.00184421, 2.81022$\times10^{-6}$, 0.00548093,  and 0.0253279. These probabilities are much smaller than the $p$-values in Tab.~\ref{tab:Pvaluesandthings} and far too small for this explanation to be reasonable. 

We next consider the explanation that any set of random data would give results that could be interpreted as being close to independent of the starting genes. To test this, we perform the following steps.

1. We randomly select a list of 6 data points out of the 24 total data points that are plotted in Tab.~\ref{fig:TemperatureRatios}. Instead of comparing each data point to others of the same temperature ratios, a random set is selected.

2. An $x$ is found that minimizes $\chi^2(x)$ for these randomly selected data points and a $p$-value is calculated for this data set in exactly the same way that the actual $\chi^2$ and $p$-values are calculated.

3. Steps 1. and 2. are repeated 1000 times and the $p$-values are averaged.

 We find the average $p$-value is 0.012. This value is smaller than any of the $p$-values given in Tab.~\ref{tab:Pvaluesandthings}. This result indicates that the differences in temperatures are not merely random fluctuations. Instead, they are true differences for the various temperatures. However, it should be noted that this $p$-value is quite close to the $p$-value for the ratio $T_3/T_1$. This indicates that the differences in temperatures are somewhat subtle.

With these two tests, we can conclude that the ratios of temperatures are close to independent of the starting genes and the small variation from independence can be explained by changes in the genes during evolution. These results are completely consistent with the work of Refs.~\cite{ToE,Vanchurin_2022}.

\section{Conclusion}
\label{sec:Conclusion}

In this work, we performed computational simulations of organisms evolving in response to an environment. The goal was to test a claim in the paper Thermodynamics of Evolution and the Origin of Life. The tested work defined an evolutionary partition function and claimed that ratios of these partition functions do not depend on starting genes. Results of the computational simulations are exactly consistent with the predictions of the work we test. Tests have been performed to falsify obvious alternative explanations. Although further investigation is warranted, these results provide strong evidence in support of the framework of The Thermodynamics of Evolution.

%%%%%%%%%%%%%%%%%%%%%%%%%%%%%%%%%%%%%%%%%%%%%%%%%%%%%%%%%
\bibliography{main}

\appendix
\begin{onecolumngrid}

\section{Simulation and Data Analysis Details}
\label{sec:SimulationDetails}

In this appendix we explain the  technicalities of the simulation. Tab.~\ref{tab:genetable} lists the starting parameters for the 6 gene combinations used. The organism follow the following rules for survival: 
Organisms' survival is determined by their energy. If an organism's energy drops below zero, the organism is removed from the simulation. One way that organisms gain energy is by passing over food which is randomly generated.  If an organism consumes a unit of food, its energy is increased by 5 and the food is removed.

Every time step, each organism loses energy according to the following formula:

$$\Delta E=-0.015r^3-0.001v^2-0.03,$$ where $E$ is the energy, $v$ is the velocity and $r$ is the size. The $v^2$ term is inspired by the kinetic energy for motion, the $r^3$ term is inspired by volume-proportional metabolism, and the constant term is inspired by fixed energy processes. However, these terms are not intended as an exact model. Instead, the rate of energy loss should be understood as a function chosen to create an emergent environment rather than a representation of any specific system. The $v$ and $r$ terms in the energy loss mean that there is an upper bound, (that varies based on the environment) for the size and maximum velocity genes.

Every time step, the position of each organism increases in proportion to its velocity. The velocity increases in proportion to its acceleration, however, the velocity may not exceed a magnitude determined by the maximum speed given by the organism's genes. The force, as mentioned in the main text, is determined by Boid motion. The genes giving separation distance, neighbor radius, alignment weight, cohesion weight, and separation weight are the parameters for this motion. In addition to this force, a force in a random direction is added with a magnitude given by a uniform distribution between zero and the gene for random force. This force is applied at an angle rotated by an angle rotated in either direction from the angle of velocity. This angle is determined randomly from a uniform distribution between 0 and the gene for random steering.

If an organism's energy is greater than 2, it has a chance to reproduce. For each organism, a uniform random number is generated between 0 and 1. If that number is less than the reproduction rate, the organism reproduces. If it does reproduce, the newly created organism and the old organism will split the energy of the old organism. The new organism will have a set of genes that is either identical to or mutated from the old organism's genes. A random number between 0 and 1 is generated. If this number is less than the value for mutation rate, the organism will mutate. If it does, the genes is incremented by the product of the old organism's mutation degree gene, the old organisms gene, and a random number between -0.01 and 0.01. 

If an organism passes over another organism, it has a chance to consume that organism. A uniform random number between 1 and 4 is generated. If that number is greater than the ratio of the sizes of the organisms, the organism is consumed. In that case, the consumed organism is removed and its energy goes to the consuming organism.

Using the simulation described here the data is generated. It is then analyzed as described in the main text.

\begin{table}
\begin{tabular}{|l | c | c | c | c | c | c |}
\hline

 & $\textbf{q}_1$ & $\textbf{q}_2$ & $\textbf{q}_3$ & $\textbf{q}_4$ & $\textbf{q}_5$ & $\textbf{q}_6$ \\
\hline

maximum speed &14.55&15.44&20.36&17.63&0.3903&12.02\\

maximum force &0.7832&0.4027&0.1234&0.9971&0.5124&0.4146\\

separation distance &50.72&121.6&18.78&75.32&48.39&41.62\\

neighbor radius &91.82&140.5&48.37&161.2&92.92&121.7\\

alignment weight &0.1000&0.02586&0.09101&0.0582&0.1046&0.071113\\

cohesion weight &0.04432&0.0596&0.04048&0.1087&0.0859&0.07735\\

separation weight &0.06947&0.1185&0.1044&0.1018&0.0766&0.1287\\

reproduction rate &0.1284&0.1935&0.08321&0.2025&0.1389&0.3327\\

mutation rate &0.9022&0.2672&0.2201&0.1227&0.5858&0.6683\\

mutation degree &0.6777&0.6098&0.8496&0.6604&0.7929&0.6406\\

random force &5.506&2.903&3.128&10.37&2.798&8.236\\

random steering &0.2907&0.09617&0.9360&0.4094&0.7702&0.0686\\

size &0.7853&0.7688&0.6687&0.5105&0.9020&0.4186\\

\hline

\end{tabular}

\label{tab:genetable}
\caption{List of initial gene values for each gene set $\textbf{q}_i$.}
\end{table}

\end{onecolumngrid}
\end{document}